\documentstyle[aps,psfig,epsf,epsfig]{revtex}
\begin{document}
\draft
\title{Low energy atomic collision with dipole interactions}
\author{B. Deb and L. You}
\address{School of Physics,
Georgia Institute of Technology
Atlanta, GA 30332, USA}
\date{\today}
\maketitle

\begin{abstract}
We apply quantum defect theory to study low energy ground state
atomic collisions including aligned dipole interactions such as
those induced by an electric field. Our results show that coupled
even ($l$) relative orbital angular momentum partial wave channels
exhibit shape resonance structures while odd ($l$) channels do
not. We analyze and interpret these resonances within the
framework of multichannel quantum defect theory (MQDT).
\end{abstract}

\pacs{34.50.-s, 34.50.Cf, 05.30.Jp, 05.30.Fk}


\section{INTRODUCTION}
Due to the tremendous progress made in laser cooling and trapping \cite{chu},
cold atomic collision has become a frontier area of research in recent
years.  The atomic Bose-Einstein condensation demonstrations \cite{bec} have
furthered this new trend of research in atomic physics.  Recently,
a topic of particular interest is the manipulation of BEC and matter-
waves properties by controlling atom-atom interactions.
Several groups have discussed mechanisms
for changing the scattering length of atomic collision using near
resonant lasers \cite{gora}, radio frequency  fields
\cite{verhaar}, Feschbach resonance due to a magnetic field
\cite{eite}, and shape resonance from dc electric field (dc-E)
induced dipole interactions \cite{marinescu}.

In this paper we present a detailed multichannel scattering
investigation of low energy atomic collisions with anisotropic dipole
collisions. Earlier studies of this problem \cite{marinescu,liyou}
have revealed several interesting features at low collision energies.
Of particular interests to this study is its potential applications
to modify (bosonic) atom-atom interaction strength with scattering
resonances and the generation of p-wave BCS states for a single
component fermi gas \cite{liyou}. In this article,
we develop a multichannel quantum defect theory (MQDT) using
exact asymptotic solutions for the long range interatomic potential
to analyze scattering resonance features. Our main aim is to explain
the physical origin of dc-E dipole interaction induced resonances
in the low energy limit. We also want to understand why such
resonances only occur in even ($l$) partial wave channels, but not
among odd ($l$) channels as numerically discovered
earlier \cite{marinescu}.
In a model study with two coupled channel ($l=0$ and $2$),
we attempt to clarify the results of complicated multichannel scattering
and illustrate the simple origin of these resonances.
We find these are {\it shape resonance} caused by the
coupling between different spatial spherical harmonic
scattering channels due to anisotropic dipole interaction.

This paper is organized as follows; First we briefly discuss
how an external dc-E modifies the effective interaction potential
between two neutral atoms. We then present a detailed
mathematical scattering formulation in Sec. II.
In section III, we describe
numerical techniques for the multichannel scattering calculation
developed earlier by Marinescu and You \cite{marinescu}, and
present additional illustrative results (for $^{85}$Rb in
the singlet molecular potential state).
The MQDT including dipole interaction is described in Sec. IV.
In section V, we analyze the structure of the observed shape
resonance with a simple model using only two (lowest angular
momentum $l$) coupled channels. The dc-E field dependent zero
energy bound states are discussed for the two channel model system
in Sec. VI. Finally we summarize and conclude in Sec. VII.

\section{Formulation}
The long-range interaction between two spherically
symmetric atoms in the ground state, is usually given
[in the London-van der Waals (LvW) formalism] by
\begin{eqnarray}
V_{0}(R)=-\frac{C_6}{R^6}-\frac{C_8}{R^8}-\frac{C_{10}}{R^{10}}\cdots,
\label{a1}
\end{eqnarray}
where $C_6$, $C_8$, and $C_{10}$ are the dispersion coefficients,
and $R$ is the internuclear distance. With this asymptotic potential,
zero energy scattering is described essentially only by one parameter:
the s-wave scattering length, $a_{\rm sc}$ \cite{huangb}.
In the absence of a dc-E,
the Hamiltonian of a model system of two spherically symmetric
neutral alkali atoms A and B can be written as
\begin{eqnarray}
H=H_{A}+H_{B}+V_{c}
\end{eqnarray}
where $H_{i=A,B}$ is the unperturbed Hamiltonian of atom $i$,
and $V_{c}$ is the Coulomb potential between charge
distributions of the two atoms and is given by the multipole
expansion from \cite{mm}
\begin{eqnarray}
V_{c}=\sum_{l,L=1}\frac{V_{lL}(\vec{r_{A}},\vec{r_{B}})}{R^{l+L+1}}
\end{eqnarray}
where $\vec{r_i}$ is the position vector of valence electron
of $i$-th atom  and $R$ is the separation between the
centers of mass of the two atoms. Take the spatial quantization
axis to be along the interatomic direction $\hat{R}$, the coefficient
$V_{lL}(\vec{r_{A}},\vec{r_{B}})$ is then given by
\begin{eqnarray}
V_{lL}=(-1)^{L} 4\pi (\hat{l}\hat{L})^{-1/2}
\sum_{m}{\mathbf{K}}_{lL}^{m}r_{A}^{l}r_{B}^{L}Y_{lm}
(\hat{r}_{A})Y_{L-m}(\hat{r}_{B})
\end{eqnarray}
with $\hat{l} \equiv 2l+1$, $\hat{L}\equiv 2L+1$, and
\begin{eqnarray}
{\mathbf{K}}_{lL}^{m}=\left[C_{l+L}^{l+m}C_{l+L}^{L+m}\right]^{1/2},
\end{eqnarray}
where $C_{L}^{m}={L!}/[{m!(L-m)!}]$.
Equation (1) of the long range dispersion is obtained from a
perturbation calculation of $V_{c}$ in the limit of large $R$
when exchange effects between different atoms are
negligible.

The presence of a dc-E distorts the spherical symmetry of an atom,
consequently, the long-range form of interatomic potential Eq. (\ref{a1})
is modified. In the $R\rightarrow \infty$ limit, the ground state
wave function $|S\rangle \equiv |n_{g},l=0,m=0 \rangle$ ($n_g$, $l$, and $m$
are respectively the principal, angular, and magnetic quantum numbers)
of an atom acquires a small $P$ angular momentum
component due to the electric dipole coupling with excited states.
Within the first order perturbation theory, the
perturbed ground state wave function can be written as
$|\phi\rangle = |S\rangle + {\cal E}|{\mathbf{P}}\rangle$,
where $|{\mathbf{P}}\rangle$ is the sum of all $P$ excited
states,
\begin{eqnarray}
|{\mathbf{P}}\rangle=\frac{1}{\sqrt{3}}\sum_{n,m}|n,l,m\rangle\frac{(n,1|r
|n_{g},0)(\hat{s}\cdot\hat{e}_{m}^{*})}{E_{n,1}-E_{n_{g},0}}.
\end{eqnarray}
$\hat{s}$ denotes the unit vector along the dc-E direction
and $r$ is the valance electron coordinate
(dipole operator) expanded in the tensor operator basis
$\hat e_m$ of $\vec r_i$.
The first order energy perturbation, i.e. interatomic
interaction potential can then be written as
\begin{eqnarray}
E^{(1)}&=&\langle \phi\phi|V_{c} |\phi\phi\rangle \\
  &=& \left( \langle SS|+ {\cal E} \langle S{\mathbf{P}}|+ {\cal E}^2
  \langle{\mathbf{P}}{\mathbf{P}}|\right) V_{c} \left( |SS\rangle +
  {\cal E} |S{\mathbf{P}}\rangle + {\cal E}^2
  |{\mathbf{P}}{\mathbf{P}}\rangle \right).
  \label{seq01}
\end{eqnarray}
To leading order in the dc-E amplitude $\cal E$, equation
(\ref{seq01}) generates an additional term,
\begin{eqnarray}
V_E(R)=-\frac{C_E}{R^3} P_2(\cos\theta), \label{a2}
\end{eqnarray}
to the LvW formalism of Eq. (\ref{a1}). $C_E=2{\cal
E}^2\alpha_1^A(0)\alpha_1^{B}(0)$ is the induced electric
dipole interaction coefficient with $\alpha_1^{A(B)}(0)$
the static atomic dipole polarizabilities of atom A(B).
$P_2(.)$ is the Legendre polynomial of order 2 and $\theta$ is the angle
between the directions of the electric field ($\hat s$) and the internuclear
axis ($\hat R$). As discovered earlier by Marinescu and You \cite{marinescu}
this electric field induced dipole interaction Eq. (\ref{a2}),
has a ``quasi long-range" character in the sense that it generates
a ``short-range" contribution to the effective potential of the
partial wave channel $l=0$ while it generates a ``long-range"
contribution (proportional to $-1/R^3$) for all other partial wave
channels ($l\neq 0$).

Assuming a weak dc-E, the value of the induced
dipole term [Eq. (\ref{a2})] is small in comparison to typical
atomic energy scale [{\it e.g.} 100 (kV/cm) is equivalent to
$1.94401\times10^{-5}$ (a.u.)]. Nevertheless, the
qualitatively different asymptotic behavior for the interaction
potential ({\it i.e.} $-1/R^3$) provides significant implications for
threshold behaviors of low energy collision.  We recall that for
a spherically symmetric short range potential (which vanishes
exponentially with increasing $R$), partial wave scattering
phase shift obeys the Wigner threshold law $\delta_l(k)\sim k^{2l+1}$
at low energies. For a long range potential vanishes as $-1/R^n$,
$\delta_l$ instead behaves in the limit of zero energy as
$k^{2l+1}$ if $l < (n-3)/2$ and as $k^{n-2}$ otherwise \cite{wu}.
For our problem, the complete long-range interatomic potential
is given by $V_0(R)+V_E(R)$, where $V_0(R)$ is the usual
long-range dispersion form Eq. (\ref{a1}) free of dc-E.
Different partial wave channels are now coupled because of
the anisotropic $Y_{20}(\hat R)\sim P_2(\hat R)$ in $V_E$.
Applying the rule
for low energy collision phase shifts to all diagonal terms of
the potential, we discover the interesting situation where all
partial wave phase shifts are proportional to $k$
due to the presence of induced dipole interaction $V_E$ \cite{marinescu}.

The anisotropic Eq. (\ref{a2}) couples different angular momentum
channels. The matrix elements for the scattering potential are simply
\begin{eqnarray}
V(R)\delta_{ll'}\delta_{m m'} -\frac{C_E}{R^3} \langle l m
|P_2(\cos\theta) |l^{\prime}m^{\prime}\rangle,
\end{eqnarray}
where $|lm\rangle = Y_{lm}(\hat{R})$ and $V(R)$ is the usual dc-E free
isotropic interatomic potential which reduces to $V_0(R)$ in the
large $R$ limit. Due to symmetry of the z-component of the angular
momentum, the matrix element
$\langle l m|P_{2} |l^{\prime}m^{\prime}\rangle$ vanishes if
$l+l^{\prime}$ is an odd number; A non-zero coupling only exists
if $l-l^{\prime}=\pm 2$ and $m=m^{\prime}$. Therefore even
and odd parity channels as well as different $m$
blocks are decoupled.

Because of the coupling among different partial wave channels,
the usual procedure for partial wave scattering calculation
needs to be modified to accommodate the dipole interaction $V_E$.
We assume the scattering wavefunction for $R \rightarrow \infty$
to be of the form
\begin{eqnarray}
R\Psi(\vec{R}) \sim R \exp(i\vec{k}\cdot\vec{R}) + f(\vec{k},\hat{R})
\exp(ikR)
\end{eqnarray}
where $\vec{k}$ is the incident momentum. The on-shell
elastic scattering is then described by $f(\vec{k},\hat{R})$
with the scattered momentum $\vec{k}^{'}=k\hat{R}$.
We expand the scattering amplitude $f(\vec{k},\hat{R})$ onto
the complete basis
\begin{eqnarray}
f(\vec{k},\hat{R})=\frac{4\pi}{k}\sum_{l m} T_{l m}(\vec{k})Y_{l
m}(\hat{R})
\end{eqnarray}
and apply the partial wave expansion
\begin{eqnarray}
\exp(i\vec{k}\cdot\vec{R}) = 4\pi \sum_{l,m}
i^{l}j_{l}(kR)Y_{lm}^{*}(\hat{k})Y_{l m}(\hat{R}).
\end{eqnarray}
We then obtain for $R \rightarrow \infty$
\begin{eqnarray}
R\psi_{k}(\vec{R})=\phi_{k}(\vec{R})=\frac{4\pi}{k}\sum_{l m}
i^{l} \left[Y_{lm}^{*}(\hat{k})\sin(kR-l\pi/2) + T_{l
m}(\vec{k})\exp(ikR-il\pi/2)\right]Y_{lm}(\hat{R}),
\end{eqnarray}
where we have used the asymptotic form $j_{l}(kr) \sim
\sin(kR-l\pi/2)/(kR)$. Therefore the scattering equations take the
multichannel form
\begin{eqnarray}
h_{l}\phi_{lm}(\vec{R})=\sum_{l^{\prime}m^{\prime}}i^{l^{\prime}-l}\langle
l m |V(\vec{R})|l^{\prime}m^{\prime} \rangle
\phi_{l^{\prime}m^{\prime}},
\end{eqnarray}
with
\begin{eqnarray}
h_{l}=-\frac{\hbar^2}{2\mu} \frac{d^{2}}{dR^2}+
\frac{\hbar^2}{2\mu}\frac{l(l+1)}{R^2}-E,
\label{seq}
\end{eqnarray}
and $\phi_{lm} = \langle lm|R\Psi(\vec{R})\rangle$.
$E=\hbar^2k^2/(2\mu)$ is the collision energy.
The boundary conditions are then given by
\begin{eqnarray}
\phi_{lm} \sim Y_{lm}^{*} (k) \sin(kR-l\pi/2) +
T_{lm}(\vec{k})\exp(ikR-il\pi/2).
\label{bd}
\end{eqnarray}

The above coupled multichannel scattering equation in
relative angular momentum ($l$) channels can also be
conveniently written in a matrix form
\begin{eqnarray}
\left[ (- \frac{\hbar^2}{2\mu}\frac{d^2}{dR^2} + V){\mathbf{I}} +
\frac{\hbar^2}{2\mu R^2} {\mathbf{L}}^{2} +
 {\mathbf{V}}_{E} \right] {\mathbf{\Phi}} = E {\mathbf{\Phi}},
 \label{seq2}
\end{eqnarray}
where $\mathbf{I}$ is the identity matrix,
$\mathbf{L}$ is the momentum operator and ${\mathbf{L}}^{2}$ takes
the form of a diagonal matrix:
${\mathbf{L}}^{2}={\mathbf{Diag}}[l(l+1),(l+2)(l+3), \cdots]$ with
$l=$ even or odd channels decoupled; $\mu = M/2$ is
the reduced mass for two identical atoms. The wave function
${\mathbf{\Phi}}$ is a column vector containing coupled
channel function $\phi_{lm}$, (i.e., different outgoing channels
$Y_{lm}(\hat{R})$). Since we need a complete set of linear
independent scattering solutions to determine $T_{lm}$ from its
expansion
\begin{eqnarray}
\frac{1}{k}T_{lm}(\vec{k})=\sum_{l'm'}t_{lm}^{l'm'}(k)Y_{l'm'}(\hat{k}),
\end{eqnarray}
${\mathbf{\Phi}}$ is made to be a matrix with different columns
characterizing scattering solutions for different incident
channels $Y_{lm}(\hat{k})$. The T-matrix elements,
$T_{lm}^{l'm'}$, can be extracted by imposing the boundary
condition Eq. (\ref{bd}) on the partial waves in the asymptotic region.
The total elastic cross section is given by
\begin{eqnarray}
\sigma_{\rm E(O)}=8\pi\sum_{l,l'={\rm even(odd)}}\sum_{m,m'}
|t_{lm}^{l'm'}|^2. \label{a5}
\end{eqnarray}
for E (even) and O (odd), respectively.
We recall that $t_{lm}^{l'm'}= -a_{\rm
sc}\,\delta_{l0}\delta_{l'0}\delta_{m0}\delta_{m'0}$ for
spherically symmetric collisions described by Eq. (\ref{a1})
without the dipole interaction.

As a technical note for caution, we emphasize that the dipole interaction
Eq. (\ref{a2}) is only valid for $R>\!>a_0$ ($a_{0}$ is Bohr radius).
In constructing the complete potential, it is therefore required to
be smoothly added to the isotropic potential, whose short range part $V_{\rm sh}$
also needs to be smoothly connected with its asymptotic form $V_0$.
We used different cut-off radius $R_{c}$ with the same
matching function for this purpose
\begin{eqnarray}
f_{c}(R)=\theta(R-R_{c}) + \theta(R_{c} - R) \exp[-(R_{c}/R-1)^2],
\end{eqnarray}
where $\theta(.)$ is the step function. We also note as elucidated
by Jackson \cite{jackson}, an additional contact term is needed for
the point dipole interaction to comply with the Gauss's law.
For our case of Eq. (\ref{a2}), this  amounts to a contact term
$-8\pi\alpha^{(A)}\alpha^{(B)}{\cal E}^2\delta(\vec R)/3$.
Such an addition is absent in our formulation since its effect is
already implicitly included in the short range potential of
$V(R)$.

\section{Results and Discussions}
The standard numerov technique was adopted in
numerically propagating Eq. (\ref{seq2}) to obtain asymptotic scattering
solutions. At lower energies, details of the potential functions
$V(R)$ becomes important. Accurate potentials were obtained from
other groups \cite{greene}. Most often we adopted a
parametrization approach to adjust a reasonable $V_{\rm sh}$
potential such that a correct $a_{\rm sc}$ was obtained.
To be able to solve the problem efficiently, we adopted
a variable step size $h$ along $R$.
With numerov we propagate from $R=0$ to asymptotic region $R_{\infty}$,
we specify the initial conditions as ${\mathbf{\Phi}}(0)=0$
and ${\mathbf{\Phi}}(h)={\mathbf{\Phi}}_{\rm arb}$, where
${\mathbf{\Phi}}_{\rm arb}$ is an arbitrary symmetric matrix
whose columns are chosen to be linearly independent vectors.
In the absence of analytical solutions, there seems
to be no clear-cut criteria for choosing the asymptotic region.
With our numerical code, in the sub-$\mu K$ energy region, typically
convergent results were only obtained for $R\ge 10^{7} a_{0}$.
In order to maintain orthogonality and
linear independence of different solution vectors during
propagation, the subspace rotation technique is performed regularly.
In the asymptotic region,
we determine (by matching to potential free motion states)
\begin{eqnarray}
\phi_{lm} = c_{1}\sin(kR-l\pi/2) + c_{2}\cos(kR-l\pi/2),
\end{eqnarray}
and construct the S-matrix with its element
given by $(c_{1}-ic_{2})^{-1}(c_{1}+ic_{2})$
and the T-matrix $({{\mathbf{S}}-1})/{(2i)}$.

In the illustrative study to be reported below, unless
stated otherwise, we always use
$^{85}$Rb atoms in its singlet potential curve.
We have employed a maximum angular quantum momentum
$l=8$ (9) for even (odd) channels, respectively.
A reasonably good
asymptotic radius $R_{\infty}$ is chosen to satisfy
$|V-C_{3}/R^3|<10^{-6} \times |{\hbar^2 l(l+1)}{R^2}/{(2\mu)} -
E|$ with $l=2$ (1) for even (odd) partial wave channels.

\begin{figure}
\psfig{file=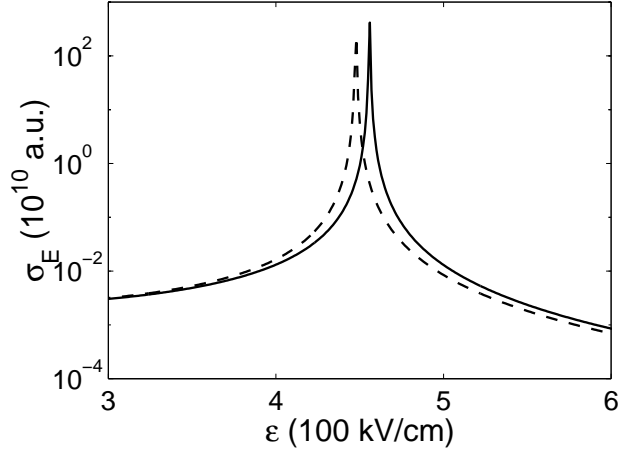,width=3.25in}
\caption{The total scattering
cross section as a function of dc-E field ${\cal E}$ for
several different cut-off radius $R_{c}=10 (a_{0})$ (dashed line)
and $R_{c}=27 (a_{0})$ (solid line). Typically an increase of $R_c$
results in a shift to higher ${\cal E}$ value of the resonance
peak at ${\cal E}_r$. Collision energy is $E=0.01$ (nk).}
\label{fig1}
\end{figure}

\begin{figure}
\psfig{file=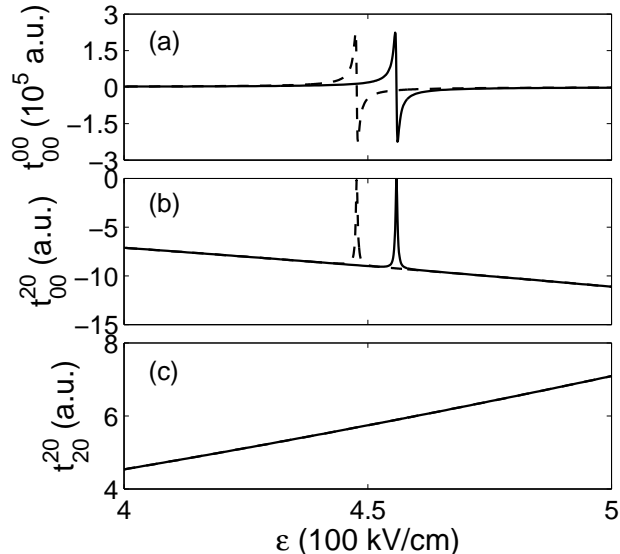,width=3.25in}
\caption{Selected reduced T-matrix elements under the same conditions
as in Fig. \ref{fig1}. In (c), the two curves are indistinguishable.}
\label{fig2}
\end{figure}

\begin{figure}
\psfig{file=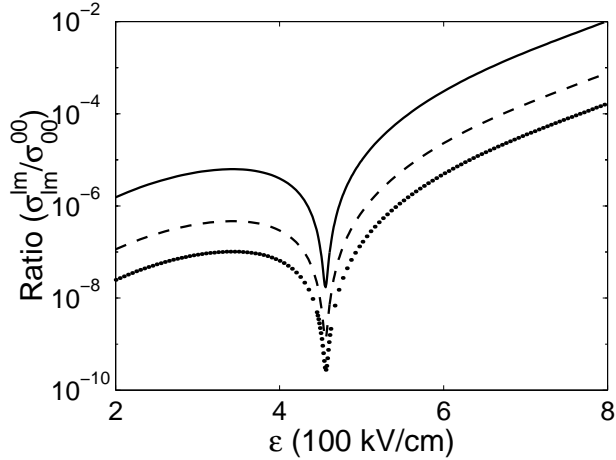,width=3.25in}
\caption{The high $l$ channel partial wave scattering cross
sections are compared to the s-wave cross section in terms of
the ratios $\sigma_{20}^{20}/\sigma_{00}^{00}$ (solid line),
$\sigma_{30}^{30}/\sigma_{00}^{00}$ (dashed line),
and  $\sigma_{40}^{40}/\sigma_{00}^{00}$ (dotted line) at
$E=0.01$ (nk).
}
\label{fig3}
\end{figure}

First, we discuss  the choice of cut-off radius $R_{c}$ and the
asymptotic radius $R_{\infty}$ for our numerical calculations.
Figures \ref{fig1}, \ref{fig2}, and \ref{fig3} illustrate effects of
different values for electric dipole cut off radius $R_{c}$.
In general, we find that
resonance peak shifts towards higher ${\cal E}$ values as $R_c$
increases. Because of the perturbative nature for the effective
long range dipole interaction, $R_c$ is always taken to be much
larger than $a_0$. Unless otherwise stated, we will use $R_c=27 (a_0)$
to present our results. The small $R_c$ dependence can be
operationally fixed by a normalization against any dc-E field
dependent experimental data.

Figure \ref{fig2} shows that, among all partial wave channels,
scattering in the s-wave ($l=0$) channel is the most sensitive
with the choice of cut-off radius, while higher $l$ channel
results remain almost unaffected.
Physically this $R_c$ dependence can be explained in terms of the
centrifugal potential term $\propto l(l+1)/R^2$ which is absent
for the s-wave channel. Comparing with Fig. \ref{fig1}, we see
that the resonance at ${\cal E}=456$ (kV/cm) is mostly due to the
contribution from $t_{00}^{00}$ (s-wave) despite the anisotropic
nature of dc-E induced dipole interaction. The coupling of s- to d-wave
is the main reason for such a resonance as
otherwise there seems no direct dipole interaction contributions
to the s-wave channel. Our extensive calculations show that typical
values of $t_{00}^{20}$ are many orders of magnitude smaller than
$t_{00}^{00}$ [for ${\cal E}<1$ (MV/cm)]. This is explicitly shown
in Fig. \ref{fig3}, where the ratio for different partial wave
cross sections are compared with the s-wave one. We see that at
low energies, scattering cross section for $l > 2$ channels
are smaller than that for the s-wave ($l=0$) by at least two
orders of magnitude for ${\cal E}<1$ (MV/cm). At and near the resonance,
scattering for $l\neq 0$ are absolutely negligible ($< 10^{-5}$ of
$\sigma_{00}^{00}$) in this case.

\begin{figure}
\psfig{file=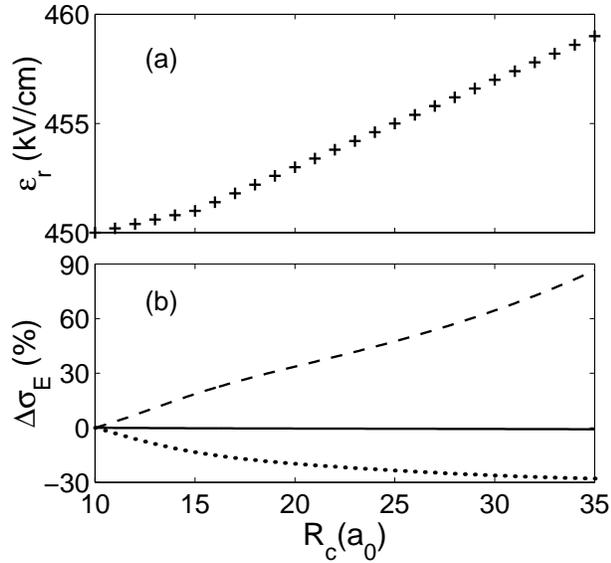,width=3.25in}
\caption{The dependence of the resonance value ${\cal E}_{r}$ (a)
and the percentage change of the total scattering cross section (b)
on $R_c$ at a selected dc-E values of ${\cal E}=200$ (kV/cm)
(solid line),
${\cal E}=500$ (kV/cm) (dashed line), and ${\cal E}=1000$ (kV/cm)
(dotted line).
$E=0.01$ (nk).}
\label{fig4}
\end{figure}

For ${\cal E} < 200$ (kV/cm), we find
the $R_c$-dependence to be marginal
as shown in Fig. \ref{fig4}. But at increased field
strength the overall effect is not always negligible ($>1$\%).

\begin{figure}
\psfig{file=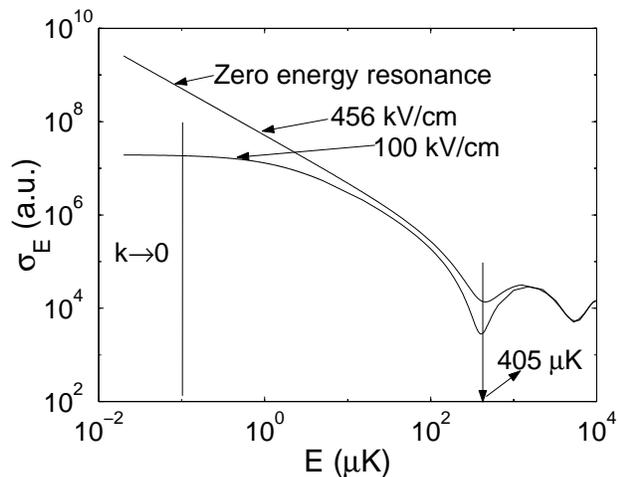,width=3.25in}
\caption{The dependence of zero energy scattering
cross section ($\sigma_E$) on collision energy for
selected dc-E values of ${\cal E}=100$ (kV/cm)
and ${\cal E}=456$ (kV/cm). At any zero energy resonance,
the scattering cross sections scales inverse
proportional to collision energy.}
\label{fig5}
\end{figure}

Next, we consider the low energy threshold dependence
of the scattering cross section. As discussed earlier,
we expect the asymptotic behaviour
$T_{lm}^{l'm'} \sim k$ at sufficiently low energies.
We have performed extensive calculations to assure that
all reported results are asymptotically converging
such that we indeed have reached this limit.
In Fig. \ref{fig5} we
display the variation of total scattering cross section $\sigma_{E}$
as a function of energy ranging from nK to mK regime.
We see that for ${\cal E}=100$ (kV/cm), $\sigma_{E}$ becomes
almost independent of energy below 1 ($\mu$K), while at
${\cal E}=456$ (kV/cm) $\sigma_{E}$
varies almost linearly with inverse of energy in
the range of $20$ (nK) to $10$ ($\mu$K). It is worth pointing out
here that at ${\cal E}=456$ (kV/cm), a resonance occurs at low
energy as discussed later resulting in a divergent low energy
cross section $\propto 1/k^2$.

\begin{figure}
\psfig{file=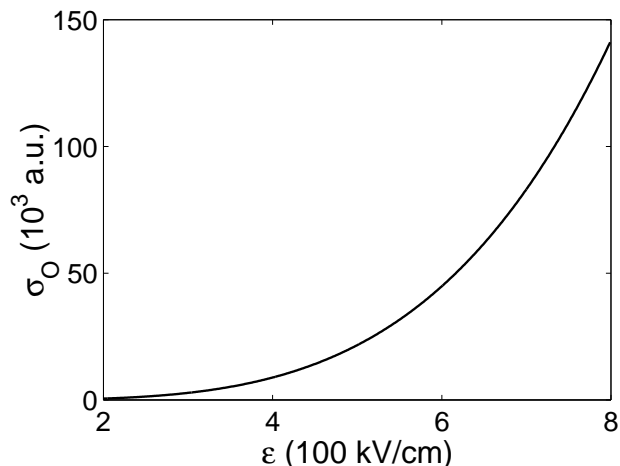,width=3.25in}
\caption{An example of the total scattering cross section
$\sigma_O$ for $^{82}$Rb (fermion)
is also found to be dominated by $t_{10}^{10}$.
No resonance occurs in this case, and
$E=0.01$ (nk). }
\label{fig6}
\end{figure}

Figure \ref{fig6} is a selected result from
odd ($l$) partial wave channels for $^{82}$Rb (fermion).
No resonance structure was ever detected in such cases.

\begin{figure}
\psfig{file=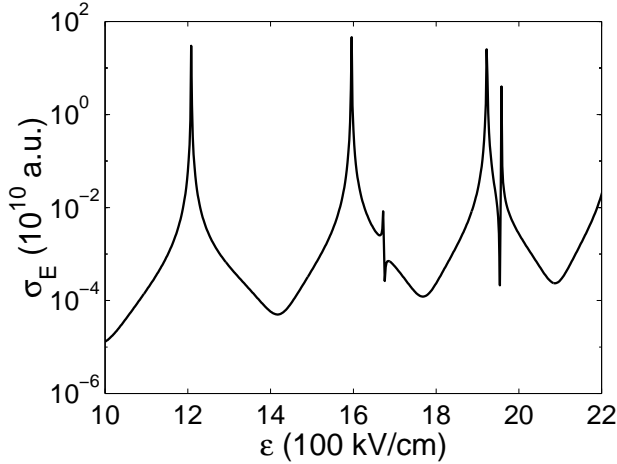,width=3.25in}
\caption{The dc-E field ${\cal E}$-dependent
total scattering cross section of $^{85}$Rb at a
collision energy $E=0.01$ (nK). Multiple resonance structures appear
in this case for ${\cal E}>1$ (MV/cm). }
\label{fig7}
\end{figure}

In Fig. \ref{fig7}, we plot the scattering cross section
of $^{85}$Rb for large fields ${\cal E}>1$ (MV/cm). As the field
strength increases the resonance become more closely spaced.
Again, the major contribution to the scattering comes
from s-wave channel, although increased coupling between
s- and higher $l$- partial waves with field strength
is responsible for the multiple peak resonance structure.

The results presented so far are obtained by radial integrations
of coupled Schr\"oedinger equations. There are certain practical
disadvantages of using such a exclusively numerical method.
First, it takes considerable amount of time to obtain the results,
since the propagation of solutions continue until the asymptotic
radius, which can be of the order of $10^7 a_{0}$ or higher, is reached.
The lower the energy, the longer the cpu time is needed because of
increased $R_\infty$; Second, the increase of either collision energy
or dc-E field strength or both calls for reduced propagation step size
which again prolongs numerical computations;
Third, as discussed earlier, the asymptotic boundary is not
well defined. Therefore, no direct physical insight is gained
about the resonance. In order to overcome these shortcomings
of the numerical method, we have developed a MQDT
for scattering with anisotropic dipole interactions.
Quantum defect theory (QDT) relies on matching numerically
integrated scattering solutions to analytic solutions
(not asymptotic plane waves). The integration
can be restricted to much shorter radius $R$ if appropriate
analytical solutions are known. This not only makes the
computation faster, the use of analytical
solutions to the long range potential also helps to gain deep insight
into resonance phenomena and low energy threshold behaviour.

\section{Quantum defect theory for anisotropic dipole interaction}
Quantum defect theory \cite{qdt} was originally formulated
to explain the spectrum of hydrogenic Rydberg atoms. In the Rydberg's
formula, $E_{n}=-R/(n-\mu)^2$ where $R$ is the Rydberg constant;
the quantum defect $\mu$ accounts for effects of the ionic core
on a highly excited electron. The idea of QDT has been successfully
applied in atomic collisions and spectroscopy over the years \cite{qdt}.
The multichannel version of QDT, known as MQDT provides a good
theoretical framework for analysis of diverse phenomena in
atomic and molecular physics. In collision theory,  MQDT requires
analytical solutions in all asymptotic channel potentials.
In the present problem including anisotropic dipole interaction,
the diagonal elements of potential matrix
goes asymptotically as $-1/R^{n}$ with $n=6$ for $l=0$ and $n=3$ for
all other channels ($l\neq 0$). The exact solutions of these power
law potentials have only recently become available through
application of secular perturbation method \cite{bogao,cavagnero}.

We first consider $n$ channels, as noted earlier, of even ($l$)
angular momentum scattering states.  In the
asymptotic region, the diagonal potential term
for s-wave channel $l=0$ goes as $-1/R^6$ while the diagonal terms
for $l>0$ channels varies as $-1/R^{3}$.
Applying MQDT, we first numerically integrate the multi-channel scattering
equation (\ref{seq2}) from $R=0$ to a certain $R=R_{0}$ such that
$C_{6}/R_{0}^{6}<\!<|\langle 20 |V_{E}(R_{0})|20\rangle|$. How to
make a judicious choice of $R_{0}$ will be discussed later. Next,
as a first approximation, we neglect off-diagonal potential
terms for $R>R_{0}$. We note more complicated procedure exists
that can incorporate off-diagonal effects in the long-range
regime ($R>R_{0}$) \cite{fano}. It will be discussed later.
The exact solutions of
$-1/R^{6}$ and $-1/R^{3}$ potentials are then matched to
numerically integrated multi-channel wave functions at $R_{0}$
in the spirit of MQDT
\begin{eqnarray}
{\mathbf{\Phi}} = {\mathbf{I\,F}} + {\mathbf{J\,G}},
\label{eqdt}
 \end{eqnarray}
 where
\begin{eqnarray}
{\mathbf{F}} && = {\mathbf{Diag}}[ f^{(6)}_{l=0},
f^{(3)}_{l=2}, \cdots f^{(3)}_{l=2(n-1)}],\nonumber\\
{\mathbf{G}} && = {\mathbf{Diag}}[ g^{(6)}_{l=0},
g^{(3)}_{l=2} \cdots g^{(3)}_{l=2(n-1)}],
\end{eqnarray}
are two diagonal matrices with $f^{(n)}_{l}$ and $g^{(n)}_{l}$
two suitably chosen linear independent base functions for $-C_{n}/R^{n}$
type potential. One can define a characteristic length scale for such a
power law potential $\beta_{n}=(2\mu C_{n}/\hbar^2)^{1/(n-2)}$.
Explicit expressions for $f_{l}^{n}$ and
$g_{l}^{n}$ are reproduced in appendix-A. Their asymptotic forms
are given in appendix-B, which for $E>0$ are grouped as
\begin{eqnarray}
 \left( \begin{array}{cc} f_{l}^{(n)}(R\rightarrow \infty)
\\ g_{l}^{(n)}(R\rightarrow \infty) \end{array} \right)
=\left( \begin{array}{cc} Z_{fb}^{(nl)} &
Z_{fc}^{(nl)}
\\ Z_{gb}^{(nl)} & Z_{gc}^{(nl)} \end{array} \right)
 \left( \begin{array}{cc} \sin(kR-\frac{l\pi}{2})
\\ \cos(kR-\frac{l\pi}{2}) \end{array} \right).
\label{as26}
\end{eqnarray}

The coefficient matrices $\mathbf{I}$ and $\mathbf{J}$ are
determined by use of the Wronskians:
${\mathbf{I}}=W({\mathbf{\Phi}},{\mathbf{G}}){\pi}/{2}$ and
${\mathbf{J}}=-W({\mathbf{\Phi}},{\mathbf{F}}){\pi}/{2}$
at $R=R_{0}$. As a convention, we set
the constant Wronskian $W(f_l^{(n)},g_l^{(n)})$ for the linear
independent base pairs $f_l^{(n)}$ and $g_l^{(n)}$ to $2/\pi$.
Substituting for the asymptotic form Eq. (\ref{as26}), we arrive at
\begin{eqnarray}
{\mathbf{F}}(R\rightarrow \infty) &&= {\mathbf{Z}}_{FB}{\mathbf{B}}
+ {\mathbf{Z}}_{FC}{\mathbf{C}},\nonumber\\
{\mathbf{G}}(R\rightarrow \infty) &&= {\mathbf{Z}}_{GB}{\mathbf{B}} +
{\mathbf{Z}}_{GC}{\mathbf{C}},
\end{eqnarray}
where ${\mathbf{Z}}_{FB(C)}$ are $n\times n$ diagonal matrices
of the form ${\mathbf{Diag}}
[Z_{fb(c)}^{(60)},Z_{fb(c)}^{(32)}, \cdots ]$ and
similarly expressions for ${\mathbf{Z}}_{GB(C)}$
with the subscript $F$ ($f$) replaced by $G$ ($g$).
${\mathbf{B}}=
{\mathbf{Diag}}[\sin(kR),\cdots,\sin(kR-{l\pi}/{2}), \cdots]$
and ${\mathbf{C}}=
{\mathbf{Diag}}[\cos(kR),\cdots,\cos(kR-{l\pi}/{2}), \cdots]$
are also diagonal matrices. Substituting these
expressions into Eq. (\ref{eqdt}), we obtain
\begin{eqnarray}
{\mathbf{\Phi}}(R\rightarrow \infty)
&&=({\mathbf{I}}\,{\mathbf{Z}}_{FB} +
{\mathbf{J}}\,{\mathbf{Z}}_{GB}){\mathbf{B}}\nonumber\\
&&+({\mathbf{I}}\,{\mathbf{Z}}_{FC} +
{\mathbf{J}}\,{\mathbf{Z}}_{GC}){\mathbf{C}}.
\end{eqnarray}
From which we find the scattering K-matrix as
\begin{eqnarray}
{\mathbf{K}} = ({\mathbf{Z}}_{FB} +
{\mathbf{K}}^{0}{\mathbf{Z}}_{GB})^{-1} ({\mathbf{Z}}_{FC} +
{\mathbf{K}}^{0}{\mathbf{Z}}_{GC}),
\end{eqnarray}
where ${\mathbf{K}}^{0}={\mathbf{I}}^{-1}{\mathbf{J}}$ is
a matrix that hopefully will depend only on shorter range
interactions ($R\leq R_{0}$). In general, ${\mathbf{K}}^{0}$
will have some $R$ dependence. For the case of a
single channel, it becomes a slowly varying function of $R$
and approaches a constant at $R_0$ where the potential
attains its power law asymptotic form ($-1/R^{n}$).
The matching point $R_{0}$ is therefore appropriately chosen
such that ${\mathbf{K}}^{0}$ becomes independent of $R$.
This is always possible as all analytic dependence on potential
and collision energy is taken care of by the various ${\bf Z}$ functions.
For multi-channel with anisotropic interactions as in the present context,
the situation become more complicated, and will be
discussed in the next section.

The S-matrix is obtained from K-matrix according to ${\mathbf{S}} =
({\mathbf{1}}- i {\mathbf{K}})^{-1}({\mathbf{1}}+
i{\mathbf{K}})$. The eigen phase shifts can be directly calculated
from the diagonalized S-matrix
${\mathbf{\Lambda}}^{-1}{\mathbf{S}}{\mathbf{\Lambda}}=
{\mathbf{Diag}}[\exp(2i\delta_{0}),\exp(2i\delta_{2}),\cdots]$,
where $\mathbf{\Lambda}$ is the unitary transformation matrix.
Thus the phase shifts for different incoming
and outgoing channels become available.

For odd ($l$) coupled partial wave channels, essentially the
same mathematical structure as the above formalism remains,
except now all diagonal potential terms are $\propto -1/R^{3}$
at large $R$. Therefore we only need analytical channel
solutions for $-1/R^3$ type potential. This makes the odd angular
momentum channel problem qualitatively different (simpler).

Before developing a model two channel problem, we briefly
summarize a technique that would allow for inclusion of
off-diagonal potential terms for $R>R_{0}$ within our
approach \cite{fano}.
In the presence of off diagonal couplings at large
interatomic separation $(R>R_{0})$, the wavefunction for
each channel $l$ can be expressed as
\begin{eqnarray}
\phi_{l} = u_{l}(R)f_{l}(R) - v_{l}(R)g_{l}(R),
\label{o32}
\end{eqnarray}
where $f_{l}$ and $g_{l}$ are the two linearly
independent base functions as defined earlier
satisfying the Schr\"odinger equation
for a $-1/R^{n}$ type potential. We will suppress the $m$ quantum
number index since even anistropic dipole potential $V_E$ remains
diagonal in $m$. $u_{l}$ and $v_{l}$ are two slowly
varying functions of $R$ defined by equations
\begin{eqnarray}
\frac{d \phi_{l}}{d R} &=& u_{l}(R) \frac{d f_{l}(R)}{d R} -
v_{l}(R)\frac{d g_{l}(R)}{d R},
\end{eqnarray}
and
\begin{eqnarray}
 \frac{d u_{l}}{d R}&=& \xi_{l}(R)  g_{l}(R),\nonumber\\
 \frac{d v_{l}}{d R} &=& \xi_{l}(R) f_{l}(R).
 \label{euv}
\end{eqnarray}
By substituting the above expressions into the
Schroedinger equation (\ref{seq}), we obtain
\begin{eqnarray}
-W(f_{l},g_{l}) \xi_{l}(R) = \frac{2\mu}{\hbar^2}\sum_{l'\neq
l}(i)^{l'-l}(V_{E})_{l l^{\prime}}\phi_{l^{\prime} }(R)
\end{eqnarray}
where $W(f_{l},g_{l})$ is the Wronskian of the base pair $f_{l}$
and $g_{l}$. The solutions of Eq. (\ref{euv}) can then be
written as
\begin{eqnarray}
u_{l}(R) &=& u_{l}(R_{0}) + \frac{\mu\pi}{\hbar^2}
\int_{R_{0}}^{R} dR^{\prime} g_{l}(R^{\prime})
\sum_{l^{\prime}\neq l}(i)^{l'-l}(V_{E})_{l
l^{\prime}}\phi_{l^{\prime}}(R'),\nonumber\\ v_{l}(R) &=&
v_{l}(R_{0}) + \frac{\mu\pi}{\hbar^2} \int_{R_{0}}^{R} dR^{\prime}
f_{l}(R^{\prime}) \sum_{l^{\prime}\neq
l}(i)^{l'-l}(V_{E})_{ll'}\phi_{l^{\prime}}(R'),
\end{eqnarray}
a form that allows for direct perturbation analysis \cite{qdt}.
The scattering matrix can be determined by evaluating values of
$u_{l}$ and $v_{l}$ at large $R$ values, since the asymptotic
expressions for $f_{l} $ and $g_{l}$ as given by Eq. (\ref{as26})
are known. The values of $u_{ l}(R_{0})$ and $ v_{ l}(R_{0})$
are given by the condition that at $R=R_{0}$ the $l$-channel wave
function $\phi_{l}$ given by Eq. (\ref{o32}) should coincide with
the corresponding wave function given by Eq. (\ref{eqdt}). Thus
$u_{ l}(R_{0})$ and $ v_{ l}(R_{0})$ are then respectively elements
of the matrices ${\mathbf{I}}$ and $-{\mathbf{J}}$.
We will, however, not pursue such a complicated calculation as it
turns out that most of the interesting physics in low energy
dipole collision can be obtained through a simpler model
calculation involving only two channels.

\section{two coupled channels}
\begin{figure}
\psfig{file=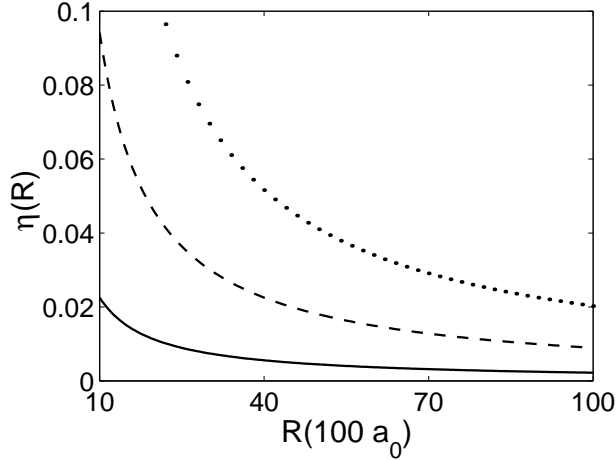,width=3.25in}
\caption{The strength of anisotropy versus $R$ for
dc-E values of ${\cal E}=100$ (kV/cm) (solid line),  ${\cal E}=1000$
(kV/cm) (dashed line), and ${\cal E}=1500$ (kV/cm) (dotted line).}
\label{fig8}
\end{figure}

To illustrate the physics of dc-E induced shape resonance,
we apply the MQDT method as outlined above, to a closed
system of two coupled channels ($l=0$ and $2$).
Given the fact that scattering for $l>2$ channels are almost negligible
at low energies as illustrated in Fig. \ref{fig3} for ${\cal E}$ field
range of interests to us, this $2\times 2$ coupled-channel system
represents a well justified model. From Eq. (\ref{seq2}) we have,
\begin{eqnarray}
\left[ \left(- \frac{\hbar^2}{2\mu}\frac{d^2}{dR^2} + V - E\right)
\left( \begin{array}{cc} 1 & 0
\\ 0 & 1 \end{array} \right) +  \frac{\hbar^2}{2\mu R^{2}}\left( \begin{array}{cc} 0 & 0
\\ 0 & 6  \end{array} \right) - \frac{C_{E}}{R^{3}}\left( \begin{array}{cc} 0 &
C_{20}
\\ C_{20} & C_{22}  \end{array} \right)
\right] \left( \begin{array}{cc} \phi_{00}
\\ \phi_{20} \end{array} \right) =0,
\end{eqnarray}
where $C_{ll'}=\langle l0|P_{2}(.)|l'0\rangle$.
As $R\rightarrow\infty$, the diagonal potential terms in the channels $l=0$ and
$l=2$  become $V_{0}= -C_{6}/R^{6}$ and $V_{2}\simeq
{6\hbar^2}/{(2\mu R^{2})}-C_{E}C_{22}/R^{3}$, respectively.
We, therefore, employ exact solutions of $-1/R^{6}$ and $-1/R^3$
potentials as obtained by B. Gao \cite{bogao} to match
numerically computed ones at a radius $R_{0}$. We first choose $R_{0}$
to satisfy condition $C_{6}/R_{0}^{6}<\!<C_{E}C_{22}/R_{0}^{3}$, i.e.,
$R_{0}>\!>[C_{6}/(C_{E}C_{22})]^{1/3}$. Therefore, as discussed in the previous
section, we expect the diagonal element of ${\mathbf{K}}^{0}$
in the $l=2$ channel, i.e., $K^{0}_{22}$ should approach a constant as
$R\rightarrow R_{0}$, provided the strength of the long range anisotropy
is negligible for $R>R_{0}$. A good measure of the relative strength of
the anisotropy can be defined according to
\begin{eqnarray}
\eta(R) =\left|\frac{V_{12}}{V_{2}-V_{0}}\right|.
\end{eqnarray}
We neglect anisotropic effects for $R>R_{0}$ if
$\eta(R>R_{0})$ drops well below unity (typically at $5 \times
10^{-3}$). In Fig. \ref{fig8} we have plotted $\eta$ as a
function of $R$ for several different values of ${\cal E}$ field.

\begin{figure}
\psfig{file=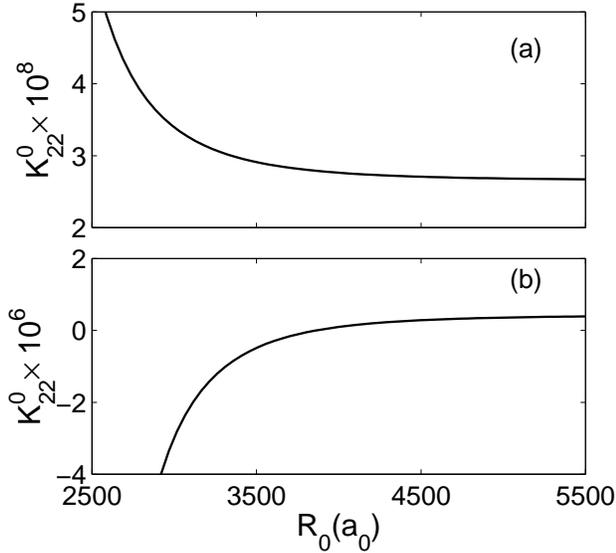,width=3.25in}
\caption{The variation of short-range K-matrix element $K_{22}^{0}$
as a function of $R$ for dc-E values of ${\cal E} = 500$ (kV/cm) (a)
and ${\cal E} = 1000$ (kV/cm) (b) at a collision energy $E=1$ (nK).}
\label{fig9}
 \end{figure}

\begin{figure}
\psfig{file=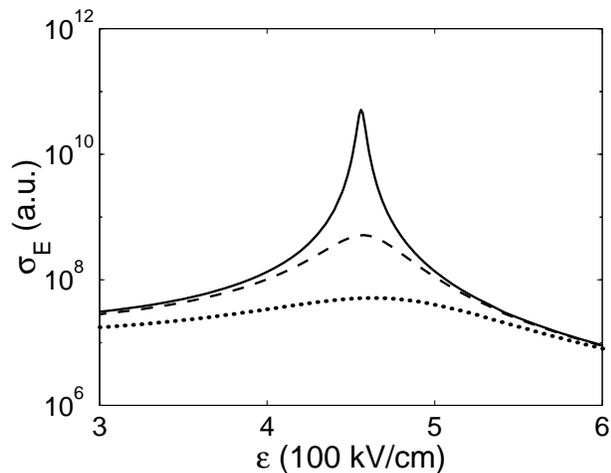,width=3.25in}
\caption{The dc-E field dependent scattering cross section $\sigma_{E}$
at three different collision energies $E=1$ (nK) (solid line), $E=100$ (nK)
(dashed line), and $E=1$ ($\mu$K) (dotted line). These results are
obtained by the MQDT method for two coupled channels.}
\label{fig10}
\end{figure}

In Figure \ref{fig9}, we display variations of $K^{0}_{22}$ as a function
of $R_0$ for two different electric field strength at $1$ (nK) collision
energy. We note that $K^{0}_{22}$ approaches a constant for $R\ge 4000 (a_0)$.
The same behavior is true at even higher ${\cal E}$ fields,
albeit, at still larger values of $R_0$.
Therefore for ${\cal E}\le 1000$ (kV/cm), we expect little loss of accuracy
by neglecting off-diagonal terms when $R\ge 4000 (a_0)$. Based on the
MQDT formulation discussed in the previous section, we have computed the
scattering wave function as well as the corresponding
T-matrix for our model of two coupled channels.
In Figure \ref{fig10}, we show the scattering cross section as a function
of dc-E field strength ${\cal E}$ at three different collisions energies.
We note that the resonance becomes more prominent as collision energy
is lowered below ($\mu$K), indicating the presence of
a bound state or a virtual bound-state (quasi-bound)
near the zero energy threshold. This calculation illustrates that
our model indeed captures the resonance
at ${\cal E}=456$ (kV/cm) as previously discussed in Fig. \ref{fig5}.
This results are almost indistinguishable from a complete numerical
calculations done earlier.
We will discuss in more detail such zero energy resonances in the
next section.

\begin{figure}
\psfig{file=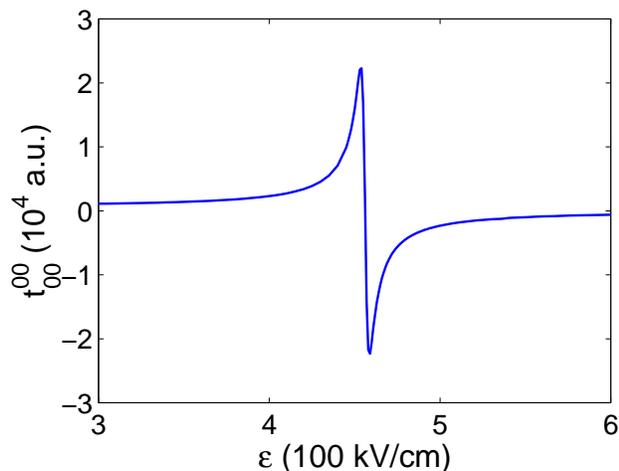,width=3.25in}
\caption{The reduced T-matrix element $t_{00}^{00}$ obtained by
MQDT for two coupled channels ($l=0,2$) at $E = 1$ (nK).
This result is in fact indistinguishable from a
numerical (two channel) calculation, although differs slightly
from a multi-channel (five channel) calculation.}
\label{fig11}
\end{figure}

As a check of consistency, we note that results obtained
from the MQDT two channel model calculation agree quite
well with those obtained by the complete numerical
calculations as long as the two channel approximation
remains valid for ${\cal E}< 1$ (MV/cm).
For ${\cal E}> 1$ (MV/cm), off-diagonal terms of the dipole
potential are not negligible for the same value of MQDT
matching radius $R_{0}$ used. A simple way of including more
effects of off-diagonal terms would be to increase $R_0$.
We therefore conclude that the MQDT does provide significant
computational advantage over the direct numerical integration technique.
It is especially efficient at lower energies when $E<10$ ($\mu$K).

\begin{figure}
\psfig{file=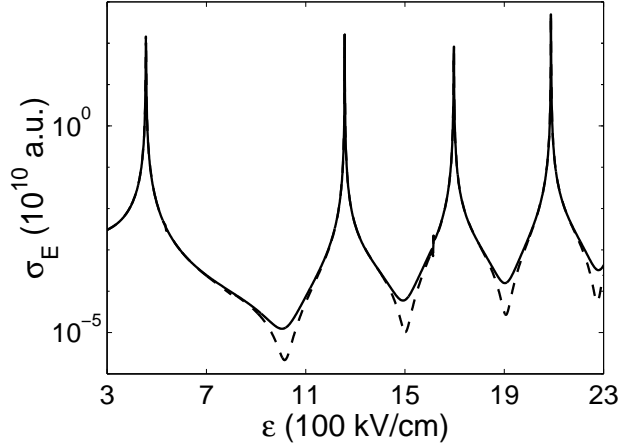,width=3.25in}
\caption{Comparison of numerical (dashed line) and
MQDT (solid line) scattering cross sections within the
two channel approximation for large dc-E fields (${\cal E} >1$ (MV/cm)
at $E = 0.01$ (nK). With such large electric fields, short range
K-matrix only become independent of $R_0$ at rather large values,
taken here to be $10000 (a_{0})$. }
\label{fig12}
\end{figure}

We now analyze the long range behaviour of our two channel model.
For the $l=0$ channel, the diagonal potential
behaves asymptotically as $-1/R^6$. This singlet potential
in fact can support many many bound states, including
those close to the zero energy threshold \cite{bogao2001}.
For $l\neq 0$ channels, on the other hand, quasi-bound states
(virtual bound states in the quasi-continuum)
may also occur due to the presence of the centrifugal potential
barrier. In the zero energy limit, scattering is
mostly due to s-wave ($l=0$) in the regime of parameters of
interest to us. Any alteration of the long-range s-wave potential
($-1/R^6$) due to electric-field-induced coupling with d-wave
leads to a modification of its bound (quasibound) spectrum near
the zero energy threshold.
The effective potential $V_{2}$ (diagonal term plus centrifugal barrier) has
a maximum (barrier) at $R=R_{m}$ given by the condition $dV_{2}/dR=0$, i.e.
$R_{m}=\mu C_{E}C_{22}/(2\hbar^2)$ and the barrier height is given
by $\delta V = \hbar^2/(\mu R_{m}^{2})$. Since coefficient
$C_{E}$ is proportional to field intensity ${\cal E}^2$,
$R_{m}$ is linearly proportional to
${\cal E}^2$. Consequently, the barrier height becomes inversely
proportional to ${\cal E}^4$. The coupling between s-
and d-wave channels may lead to combined new bound or
quasi-bound states or both at near-zero energies.
The existence of these states is manifested in the form of
scattering resonance. For instance when an incident low energy
atom from s-wave channel hits a quasi-bound states supported by the
d-wave centrifugal barrier, we may have a situation resemble
what is commonly known as a Feshach resonance \cite{eite}.
If there exists a bound state
at or near zero energy, scattering cross sections will
consequently be enhanced many-fold. This is indeed the case
found in our earlier extensive numerical calculations.
As will be proven in the next section on bound states, such
resonance structures can be fully explained based on
zero energy bound states using the MQDT.

\begin{figure}
\psfig{file=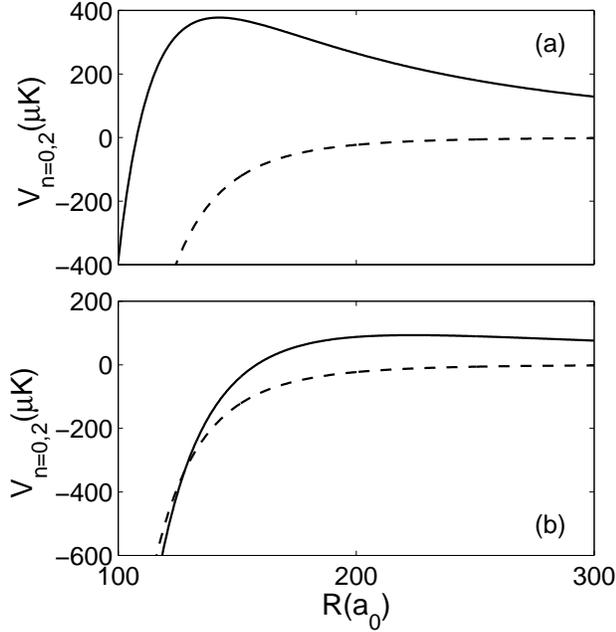,width=3.25in}
\caption{The diagonal potential terms $V_{0}$ (dashed line) for $l=0$
and $V_{2}$ (solid line) for $l=2$ at two different dc-E field
strength ${\cal E}=456$ (kV/cm) (a) and ${\cal E}=1500$ (kV/cm) (b).
The larger the ${\cal E}$ field, the further out in $R$ where
the two potential curves crosses. The height of the $V_{2}$
potential barrier decreases with increasing ${\cal E}$.}
\label{fig13}
\end{figure}

\section{Bound states}
As discussed in the previous section, low energy scattering resonance
is a signature of a bound or quasibound state near zero energy.
At some critical dc-E field strength, a new bound state is formed at
micro- or submicro-Kelvin energy leading to the observed zero-energy
shape resonance. In order to elucidate this point explicitly,
we again rely on the two-channel model and find its last bound or
quasibound state just below zero energy by the MQDT formulation.
The asymptotic form of linearly independent base pairs
$f_{l}^{(n)}$ and $g_{l}^{(n)}$ satisfying the Schroedinger equation
with a potential $-1/R^{n}$ for $E<0$ can be expressed as
\begin{eqnarray}
 \left( \begin{array}{cc} f_{l}^{(n)}(R\rightarrow \infty)
\\ g_{l}^{(n)}(R\rightarrow \infty) \end{array} \right)
=\left( \begin{array}{cc} W_{f-}^{(nl)} & W_{f+}^{(nl)}
\\ W_{g-}^{(nl)} & W_{g+}^{(nl)} \end{array} \right)
 \left( \begin{array}{cc} \exp(\kappa R)
\\ \exp(-\kappa R) \end{array} \right),
\end{eqnarray}
where $W$'s are chosen to be real functions. For the two coupled channels
discussed in the previous section, we have
\begin{eqnarray}
W_{f-}^{(60)}&=&(4\pi\kappa)^{-1/2}[(X_{0}^{2}+Y_{0}^2)\sin(\pi\nu)]^{-1},\nonumber\\
&\times&
\left[\{\beta\sin(\pi\nu)+\alpha\cos(\pi\nu)\}G(-\nu)-\alpha
G(\nu)\right],\nonumber\\
W_{g-}^{(60)}&=&(4\pi\kappa)^{-1/2}[(X_{0}^{2}+Y_{0}^2)\sin(\pi\nu)]^{-1}\nonumber\\
&\times&
\left[\{\alpha\sin(\pi\nu)-\beta\cos(\pi\nu)\}G(-\nu)+\beta
G(\nu)\right],
\end{eqnarray}
and
\begin{eqnarray}
W_{f-}^{(32)} &=&2(2\pi\kappa)^{-1/2}\left[\frac{C_{+}(-\nu)}{G(-\nu)}\Delta^{\nu}-
\frac{C_{+}(\nu)}{G(\nu)}\Delta^{-\nu}\right],\nonumber\\
W_{g-}^{(32)}&=&2(2\pi\kappa)^{-1/2}
\left[-\frac{C_{-}(-\nu)}{G(-\nu)}\Delta^{\nu}+
\frac{C_{-}(\nu)}{G(\nu)}\Delta^{-\nu}\right].
\end{eqnarray}
Except for $\hbar^2\kappa^2/(2\mu)=-E$, all
notations used here follow from earlier definitions as in appendix-A.
After some tedious algebra, the condition for bound states of
our two channel model becomes
\begin{eqnarray}
{\rm Det}[{\mathbf{Q}}+{\mathbf{K}}^{0}]=0,
\label{cbeq}
\end{eqnarray}
where ${\mathbf{Q}}$ is a diagonal matrix given by
\begin{eqnarray}
{\mathbf{Q}}={\bf Diag}[W_{f-}^{(60)},W_{f-}^{(32)}]
\left({\bf Diag}[W_{g-}^{(60)},W_{g-}^{(32)}]\right)^{-1}.
\end{eqnarray}
The bound state wave function at discrete energy $E_{i}$
can then be expressed as
\begin{eqnarray}
\Psi_{i}(R)=({\mathbf{F}}+{\mathbf{K}}^{0}{\mathbf{G}})M,
\label{bs}
\end{eqnarray}
where ${\mathbf{F}}$, ${\mathbf{G}}$, and ${\mathbf{K}}^{0}$ are
similar to matrices defined in Sec. IV, and $M$ is a column vector.
Using the asymptotic form of base pairs functions, Eq. (\ref{bs}) can be
rewritten as
\begin{eqnarray}
\Psi_{i}(R\rightarrow\infty)=\exp(-\kappa R){\mathbf{Q}}_{+}M +
\exp(\kappa R){\mathbf{Q}}_{-}M,
\end{eqnarray}
where
\begin{eqnarray}
{\mathbf{Q}}_{\pm}={\bf
Diag}[W_{f\pm}^{(nl)},W_{f\pm}^{(n'l')},\cdots]+{\mathbf{K}}^{0}
{\bf Diag}[W_{g\pm}^{(nl)},W_{g\pm}^{(n'l')},\cdots],
\label{beq}
\end{eqnarray}
The bound state energy is therefore given by the requirement that
the exponentially rising part of Eq. (\ref{beq}) vanishes, i.e.,
\begin{eqnarray}
{\rm Det}[Q_{-}]=0.
\end{eqnarray}

The matching radius $R_{0}$ in this case is chosen to be in
classically allowed region of diagonal channel potentials.
The condition for the existence of a new bound state,
i.e. Eq. (\ref{cbeq}) is then given by
\begin{eqnarray}
(Q_{11}+K^{0}_{11})(Q_{22}+K^{0}_{22})-K^{0}_{12}K^{0}_{21}=0,
\label{cbeq1}
\end{eqnarray}
where the subscript indices denote respective elements
of $ {\mathbf{Q}}$ and ${\mathbf{K}}^{0}$ matrices, and
$(1/Q_{11})$ and $Q_{22}$ are the $\chi$ functions for a
pure $-1/R^6$ or $-1/R^3$ potential
as defined in Ref. \cite{bogao}.
Since the value of $K_{12}$ is typically much smaller than
$K_{11}$ near resonance, as an approximation, we can drop the two
off-diagonal elements from ${\mathbf{K}}^{0}$. We then obtain two
uncoupled bound spectrum series, given respectively by
\begin{eqnarray}
Q_{11}+K^{0}_{11}=0,
\label{cbeq2}
\end{eqnarray}
 and
\begin{eqnarray}
 Q_{22}+K^{0}_{22}=0.
 \label{2nd}
\end{eqnarray}
The first condition Eq. (\ref{cbeq2}) gives bound states
predominantly supported by s-wave channel. The effect of
channel-coupling due to external electric field enters only
through the parameter $K^{0}_{11}$.
In this first approximation, the off-diagonal potential
terms are neglected for $R>R_{0}$. As commented earlier,
the short-range K-matrix ${\mathbf{K}}^{0}$ is a slowly
varying function of collision energies near the zero energy threshold.
We can then extrapolate their values from $E>0$ (for scattering)
to the bound state case $E_i\le 0$. In the present case of zero energy
bound states satisfying $E_i\sim 0$, we basically used the same
short range K-matrix as obtained from a converged low energy
scattering calculation.
We compute bound states near the zero energy threshold
that satisfy
\begin{eqnarray}
|E_{i}|<\!<C_{6}/R_{0}^6<C_{3}/R_{0}^3.
\end{eqnarray}
For an asymptotic $-1/R^3$ potential as in $l=2$,
the matching radius $R_{0}$ is taken to be smaller than the
characteristic length scale $\beta_{3}=2\mu C_{E}C_{22}/\hbar^2$
of the potential.

\begin{figure}
\psfig{file=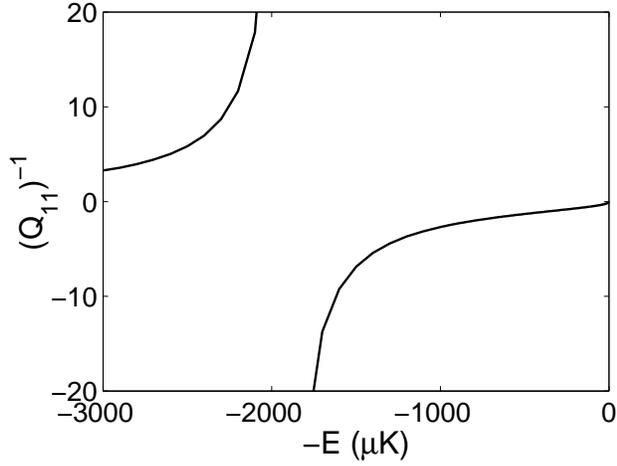,width=3.25in}
\caption{The inverse of $Q_{11}$ as a function of bound
state energy in ($\mu$K) for $^{85}$Rb. }
\label{fig14}
\end{figure}

In Figure \ref{fig14}, we show the energy dependence of $Q_{11}^{-1}$.
The last bound state of isolated s-wave $l=0$ channel (with
asymptotic potential $-1/R^6$) is then given by
crossing points of $Q_{11}$ and -$K_{11}^{0}$. In the absence
of an external electric field, the energy of the last bound state
supported by an asymptotic $-/R^6$ potential of $^{85}$Rb is
about $2$ (mK) which is in fact far beyond the zero energy limit
appropriate for the present discussion. In the zero energy limit
($l=0$),
$R_{0}$ can be taken arbitrarily large for the condition given by
by the single channel bound state Eq. (\ref{cbeq2}).

\begin{figure}
\psfig{file=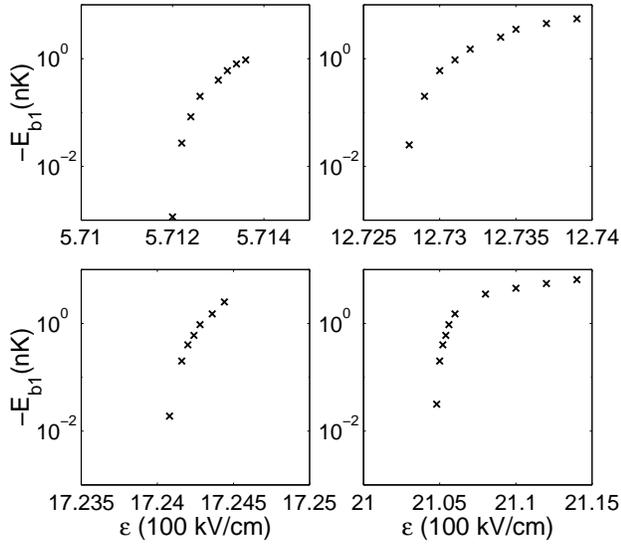,width=3.25in}
\caption{Bound state energies ($E_{b1}$) in the $l=0$ channel as
a function of electric field ${\cal E}$ in the zero energy limit.}
\label{fig15}
\end{figure}

From Fig. \ref{fig14}, we see that a new zero energy
bound state appears only when $K_{11}^{0}$ becomes infinite.
In Fig. \ref{fig15}, we display variations of the subsequent
four zero energy bound state energies as a function of
${\cal E}$. We used $K_{11}^{0}$ as determined at a positive energy
near zero energy threshold and matched analytical solutions with
the numerical ones at a relatively large $R$ ($R=4000 a_{0}$ for
${\cal E}\!<1$ MV/cm, $R=10000 a_{0}$ for ${\cal E}\!>1$ MV/cm).
It is interesting to note that these resonances agree quite
well with results obtained from a complete numerical
multichannel scattering calculation presented before in Figs.
\ref{fig1} and \ref{fig7}.

\begin{figure}
\psfig{file=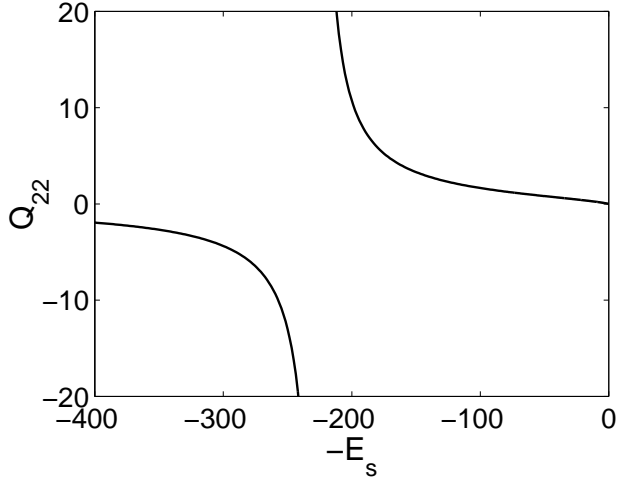,width=3.25in}
\caption{$Q_{22}$ as a function of scaled energy
$E_{s}=2\mu C_{E}C_{22}E/\hbar^2$ for $^{85}$Rb.
For an asymptotic $-1/R^3$ potential as in the $l=2$ channel,
use of this scaled energy helps to illustrate the long range
interaction. The last bound state is given by crossing
points of $Q_{22}$ with -$K^{0}$, just as in the $l=0$ s-wave case.}
\label{fig16}
\end{figure}

The second condition Eq. (\ref{2nd}) corresponds to bound
state series of the $l=2$ channel. In addition to dc-E effects
accumulated in $Q_{22}$, the asymptotic wave function is now
chosen to include the field induced diagonal term $\propto -1/R^3$.
In Figure \ref{fig16} we plot the energy and dc-E field
dependent $Q_{22}$. It resembles $Q_{11}^{-1}$ rather than
$Q_{11}$ because of a different choice between $g$ and $f$ functions.
At field values of interest to us, we find the short-range
${\bf K^{0}}$ matrix actually changes rapidly in this case
with the (relative small) matching radius $R_0$.
Therefore, in the classically allowed region, we were not able to
perform a correct MQDT with an energy and $R_0$ independent
$K_{22}^0$. Nevertheless, to explore the physics qualitatively,
we instead choose a radius $R_0=105 (a_0)$
(in the classically allowed region of
the $l=2$ diagonal potential) such that it reproduces the first bound
state as in the s-wave channel between ${\cal E}=500$ and $600$
(kV/cm). As we can see from Fig. \ref{fig15}, the third bound state
then also seem to be reproduced correctly.

\begin{figure}
\psfig{file=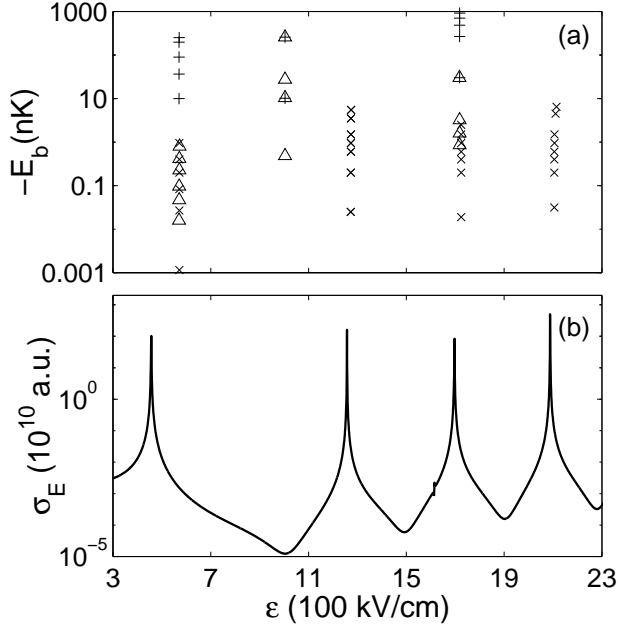,width=3.25in}
\caption{The comparison between zero energy bound states as obtained
from MQDT for a single channel $l=0$ ($\times$), $l=2$ ($+$), and
the two coupled channels ($\triangle$) in (a), and
their comparison with numerical results from the two channel
model (b). We note the s-wave single channel results matched in
the classically forbidden region at $R_0\ge 4000 (a_0)$ reproduces
exactly all numerically calculated scattering resonance.
Note that the numerical two channel results here (b) differ somewhat from
the complete multi-channel results given in Fig. \ref{fig7}.
}
\label{fig17}
\end{figure}

We now proceed to estimate the combined
bound state sequence by solving for Eq. (\ref{cbeq1}). In this
case, for bound state energy larger than a few $\mu$K, the
classically forbidden region limits the choice of
a matching radius to be within 100-150 $a_0$, which is much
smaller than the typical value of 4000 $a_0$ used for the
scattering calculations. We again choose the same
matching radius at $105 (a_{0})$ as for the single
$l=2$ channel [Eq. (\ref{2nd})] discussed above. Obviously,
there is no reason to trust the obtained results seriously
as the neglected anisotropic effects become significant at
such a small $R_0$. Nevertheless, it is interesting
to compare the obtained results as in Fig. \ref{fig17}
for these three separate approximations. Note the comparison
of bound state locations with the numerical calculated
zero energy scattering cross section from the same two channel
model. We find that the $l=0$ s-wave bound state locations
match quite well with the numerically calculated resonance
structures. This gives us the confidence to assign these
resonances as due to dc-E field induced zero energy bound states,
thus they are indeed {\it shape resonance}.
As the strength of electric field is
increased to a certain critical value, a new bound or quasibound
state appears at the zero energy threshold. Since we neglect the
off-diagonal potential terms for $R>R_{0}$, the bound state
energies calculated here only provides approximate estimates,
they differ from results reported in Fig. \ref{fig8} obtained from
a complete multichannel calculation. The fact that a single
s-wave ($l=0$) channel completely captures the resonance structure
is perhaps also not surprising, as it is indeed the dominant
channel found numerically before. By choosing the matching
radius $R_0$ to be sufficiently large, the anisotropic dipole
induced coupling with other higher $l$ channels are included
through the short range ${\bf K}^0$ matrix.

Finally we note that MQDT bound states technique used here
may also be improved with a perturbative method as discussed
earlier to include additional asymptotic ansiotropic effects \cite{fano}.

\section{Conclusion}
We have presented a detailed analysis of low energy
atomic collisions including electric field induced
anisotropic dipole interaction. We have discussed
both multichannel numerical and MQDT methods for the
scattering calculations. In particular, we have
highlighted the zero-energy resonance phenomena due
to formation of long-range bound or quasi-bound states
near the zero energy threshold. At low energies
($E \sim$ nK) and with reasonable dipole interaction strength,
the scattering cross section is predominantly
due to the s-wave for coupled even $l$ angular momentum channels.
We also note that the zero energy resonance due to formation
of new bound states only occurs in the coupled even $l$ channels,
but not in the odd $l$ channels. This is mainly due to the
presence of the s-wave potential well (without centrifugal potential)
in the coupled even $l$ blocks.
At positive energies, the s-wave channel is open throughout
the entire range ($0<R<\infty$) of its diagonal potential,
while other channels ($l>0$) have locally closed regions near zero energy.
When s-wave is present, atoms can
penetrate into shorter $R$ range via scattering and multichannel coupling
and therefore zero-energy resonance is more likely to occur.
By manipulating such scattering resonance with dc-E induced dipole
interaction, it is possible to change low energy scattering properties
such as the scattering length and total or partial wave scattering cross
sections. In contrast to the
magnetic field-induced Feshbach resonance where coupling between
different hyper-fine levels (different internal states) are
involved, electric field-induced resonance is due to coupling
among different rotational (external) states of the two
colliding atoms, and only one threshold (or internal state) is
present in the system. In the present paper, using a model
two channel MQDT calculation, we have provided a clear physical
picture of these resonances and provided analytical means for
estimating their locations. The availability of approximate
analytical forms for these near zero-energy bound states allows for
detailed examination of bound-free and free-free motional
Franck-Condon factors.

\acknowledgments We thank Mircea Marinescu for his contributions
at the earlier stages of this work.
We thank Bo Gao and F. Robicheaux for several helpful communications.
This work is
supported by the the NSF grant No. PHY-9722410.

\appendix
\section{Analytic solutions for $-1/R^{n=3,6}$-type potential}
The base pair for $-1/R^{6}$ potential is given by
\begin{eqnarray}
f_{l}^{6} &&=\frac{1}{\sqrt{2}}(\alpha^{2}+\beta^{2})^{-1}[\alpha
\tilde{f}_{l}(r)-\beta \tilde{g}_{l}(r)],\nonumber\\
g_{l}^{6} &&=\frac{1}{\sqrt{2}}(\alpha^{2}+\beta^{2})^{-1}[ \beta
\tilde{f}_{l}(r)+ \alpha \tilde{g}_{l}(r)],
\end{eqnarray}
where
\begin{eqnarray}
\tilde{f}_{l}(r) &&=\sum_{m=-\infty}^{\infty} b_{m}
r^{1/2}J_{\nu+m}\left(\frac{1}{2}(r/\beta_{6})^{-2}\right),\nonumber\\
\tilde{g}_{l}(r) &&=\sum_{m=-\infty}^{\infty} b_{m}
r^{1/2}J_{\nu+m}\left(\frac{1}{2}(r/\beta_{6})^{-2}\right),
\end{eqnarray}
with
 \begin{eqnarray}
\alpha &&=\cos[\pi(\nu-\nu_{0})/2]X_{l}-\sin[\pi(\nu-\nu_{0})/2]Y_{l},\nonumber\\
\beta &&=\sin[\pi(\nu-\nu_{0})/2]X_{l}+\cos[\pi(\nu-\nu_{0})/2]Y_{l},
\end{eqnarray}
and
 \begin{eqnarray}
X_{l} &&=\sum_{m=-\infty}^{\infty}(-1)^{m}b_{2m},\nonumber\\
Y_{l} &&=\sum_{m=-\infty}^{\infty}(-1)^{m}b_{2m+1},
\end{eqnarray}
and \cite{bogao}
\begin{eqnarray}
b_{j} &&=(-\Delta)^{j}\frac{
\Gamma(\nu)\Gamma(\nu-\nu_{0}+1)\Gamma(\nu+\nu_{0}+1)}
{\Gamma(\nu+j)\Gamma(\nu-\nu_{0}+j+1)\Gamma(\nu+\nu_{0}+j+1)}c_{j}(\nu),\nonumber\\
b_{-j} &&=(-\Delta)^{j} \frac{
\Gamma(\nu-j+1)\Gamma(\nu-\nu_{0}-j)\Gamma(\nu+\nu_{0}-j)}
{\Gamma(\nu+1)\Gamma(\nu-\nu_{0})\Gamma(\nu+\nu_{0})}c_{-j}(-\nu),
\label{bj}
\end{eqnarray}
where $j$ is a positive integer, $\Delta$ is a scaled energy,
$\Delta=k^{2}\beta_{6}^{2}/16$ with $k=2\mu E/\hbar^{2}$;
$\nu_{0}$ is related to the angular momentum $l$, $\nu_{0}
=(2l+1)/4$, and $c_{j}(\nu)=b_{0}Q(\nu)Q(\nu+1)\cdots Q(\nu+j-1)$
with $b_{0}$ being a normalization constant; and  $Q(\nu)$ is
given by a continued fraction,
\begin{eqnarray}
Q(\nu)={1 \over 1-\Delta^{2}{1 \over (\nu+1)(\nu+2)
[(\nu+1)^{2}-\nu_{0}^{2}][(\nu+2)^{2}-\nu_{0}^{2}]} Q(\nu+1)}.
\end{eqnarray}
Here $\nu$ is a root of a characteristic function
\begin{eqnarray}
\Lambda_{l}(\nu,\Delta)=(\nu^{2}-\nu_{0}^{2})-(\Delta^{2}/\nu)[\tilde{Q}(\nu)-\tilde{Q}(-\nu)],
\end{eqnarray}
where
$\tilde{Q}(\nu)=[(\nu+1)\{(\nu+1)^{2}-\nu_{0}^{2}\}^{-1}Q(\nu)$.

Next, we present the solutions of $-C_{3}/R^{3}$ potential \cite{bogao}.
The base pair can be expressed as
\begin{eqnarray}
f_{l}^{3}(r) &&=
\frac{2}{D(\nu)}\left[\frac{1}{G(-\nu)}C_{+}(\nu)\xi(r) -
\frac{1}{G(\nu)}C_{+}(-\nu)\eta(r)\right],\nonumber\\
g_{l}^{3}(r) &&=\frac{2}{D(-\nu)}\left[\frac{1}{G(-\nu)}C_{-}(\nu)\xi(r)
- \frac{1}{G(\nu)}C_{-}(-\nu)\eta(r)\right],
\end{eqnarray}
where
\begin{eqnarray}
 C_{\pm}(\nu) &&=\cos[\pi(\nu_{0}/2-\nu)] \pm \sin[\pi(\nu_{0}/2-\nu)],\nonumber\\
 D(\nu) &&=C_{+}(\nu)C_{-}(-\nu)-C_{+}(-\nu)C_{-}(\nu),\nonumber\\
 G(\nu) &&=|\tilde{\Delta}|^{-\nu}\frac{\Gamma(1+\nu_{0}+\nu)\Gamma(1-\nu_{0}+\nu)}
 {\Gamma(1-\nu)}C(\nu),
\end{eqnarray}
with $\tilde{\Delta}=k\beta_{3}/2$ and $\nu_{0}=l+1/2$. Here
$\xi(r)$ and $\eta(r)$ are two linearly independent functions:
\begin{eqnarray}
\xi(r) &&=\sum_{-\infty}^{\infty} b_{m} r^{1/2} J_{\nu+m}(kr),\nonumber\\
\eta(r) &&=\sum_{-\infty}^{\infty} (-1)^{m} b_{m} r^{1/2}
J_{-\nu-m}(kr).
\end{eqnarray}
Here $b_{j}$ is given by the similar expression as in Eq. (\ref{bj})
with $\Delta$ replaced by $\tilde{\Delta}$, and $\nu$ is a root of the
corresponding characteristic equation and
$C(\nu)=\lim_{j\rightarrow \infty}c_{j}$.

\section{Asymptotic expansion for the analytic solutions}
In this Appendix, we write the asymptotic form of a pair of
linearly independent base functions for a power law potential
($-1/R^n$). For a $-/R^6$ potential, the asymptotic behaviours of
the base pair for positive energy are given by
\begin{eqnarray}
f_{l}^{(6)}(R\rightarrow\infty) &&\sim(1/\pi
k)^{1/2}\left[Z_{fb}^{(6l)}\sin(kR-l\pi/2) +
Z_{fc}^{(6l)}\cos(kR-l\pi/2)\right],\nonumber\\
g_{l}^{(6)}(R\rightarrow\infty) &&\sim(1/\pi
k)^{1/2}\left[Z_{gb}^{(6l)}\sin(kR-l\pi/2) +
Z_{gc}^{(6l)}\cos(kR-l\pi/2)\right],
\end{eqnarray}
where
\begin{eqnarray}
Z_{fb}^{(6l)} &=&  [(X_{l}^{2}+Y_{l}^2)\sin(\pi\nu)]^{-1}
\nonumber \\ &\times&
\left[-(-1)^l\{\beta\sin(\pi\nu)+\alpha\cos(\pi\nu)\}
G(-\nu)\sin(\pi\nu-l\pi/2-\pi/4) -\alpha
G(\nu)\sin(\pi\nu-l\pi/2-\pi/4)\right],\nonumber\\
Z_{fc}^{(6l)}&=&-[(X_{l}^{2}+Y_{l}^2)\sin(\pi\nu)]^{-1}, \nonumber\\
&\times&
\left[-(-1)^l\{\beta\sin(\pi\nu)+\alpha\cos(\pi\nu)\}G(-\nu)\cos(\pi\nu-l\pi/2-\pi/4)
-\alpha G(\nu)\sin(\pi\nu-l\pi/2-\pi/4)\right],\nonumber\\
Z_{gb}^{(6l)}&=&[(X_{l}^{2}+Y_{l}^2)\sin(\pi\nu)]^{-1}, \nonumber\\
&\times&
\left[-(-1)^l\{\alpha\sin(\pi\nu)-\beta\cos(\pi\nu)\}G(-\nu)\sin(\pi\nu-l\pi/2-\pi/4)
+\beta G(\nu)\cos(\pi\nu-l\pi/2-\pi/4)\right],\nonumber\\
Z_{gc}^{(6l)}&=&-[(X_{l}^{2}+Y_{l}^2)\sin(\pi\nu)]^{-1},\nonumber\\
&\times&
\left[-(-1)^l\{\alpha\sin(\pi\nu)-\beta\cos(\pi\nu)\}G(-\nu)\cos(\pi\nu-l\pi/2-\pi/4)
+\beta G(\nu)\sin(\pi\nu-l\pi/2-\pi/4)\right].
\end{eqnarray}

Next, for a $-1/R^3$ potential, the corresponding $Z$ functions
are given by
\begin{eqnarray}
Z_{fb}^{(3l)} &&= \frac{2}{D(\nu)}
\left[\frac{1}{G(-\nu)}C_{+}(\nu)\{\cos[\pi(\nu -\nu_{0})/2]X_{l}-
\sin[\pi(\nu-\nu_{0})/2]Y_{l}\}  -
\frac{1}{G(\nu)}C_{+}(-\nu)\tilde{B}\right], \nonumber\\
Z_{fc}^{(3l)} &&= -\frac{2}{D(\nu)}
\left[\frac{1}{G(-\nu)}C_{+}(\nu)\{\sin[\pi(\nu -\nu_{0})/2]X_{l}-
\cos[\pi(\nu-\nu_{0})/2]Y_{l}\}  -
\frac{1}{G(\nu)}C_{+}(-\nu)\tilde{C}\right], \nonumber\\
Z_{gb}^{(3l)} &&= \frac{2}{D(\nu)}
\left[\frac{1}{G(-\nu)}C_{-}(\nu)\{\cos[\pi(\nu -\nu_{0})/2]X_{l}-
\sin[\pi(\nu-\nu_{0})/2]Y_{l}\}  -
\frac{1}{G(\nu)}C_{-}(-\nu)\tilde{B}\right], \nonumber\\
Z_{gc}^{(3l)} &&= -\frac{2}{D(\nu)}
\left[\frac{1}{G(-\nu)}C_{-}(\nu)\{\sin[\pi(\nu -\nu_{0})/2]X_{l}-
\cos[\pi(\nu-\nu_{0})/2]Y_{l}\}  -
\frac{1}{G(\nu)}C_{+}(-\nu)\tilde{C}\right],
\end{eqnarray}
where
\begin{eqnarray}
\tilde{B} &&=\{\cos[\pi(\nu -\nu_{0})/2]X_{l}-
\sin[\pi(\nu-\nu_{0})/2]Y_{l}\}\cos(\pi\nu)+ \{\sin[\pi(\nu
-\nu_{0})/2]X_{l}- \cos[\pi(\nu-\nu_{0})/2]Y_{l}\}\sin(\pi\nu), \nonumber\\
\tilde{C} &&=-\{\cos[\pi(\nu -\nu_{0})/2]X_{l}-
\sin[\pi(\nu-\nu_{0})/2]Y_{l}\}\sin(\pi\nu)+ \{\sin[\pi(\nu
-\nu_{0})/2]X_{l}- \cos[\pi(\nu-\nu_{0})/2]Y_{l}\}\cos(\pi\nu).
\end{eqnarray}


\begin{references}
\bibitem{chu}C. Cohen-Tannoudji, Rev. Mod. Phys. {\bf 70}, 707
(1998); S.Chu, {\it ibid} {\bf 70},685 (1998); W. D. Philips,
{\it ibid} {\bf 70}, 721 (1998).

\bibitem{bec}M. Anderson, J. R. Ensher, M. R. Matthews,
C. E. Wieman, and E. A. Cornell, Science {\bf 269}, 198 (1995); C.
C. Bradley, C. A. Sackett, J. J. Tollett, and R. G. Hulet, Phys.
Rev. Lett. {\bf 75}, 1687 (1995); C. C. Bradley, C. A. Sackett and
R. G. Hulet, {\it ibid}. {\bf 78}, 985 (1997); K. B. Davis, M. O.
Mewes, M. R. Andrews, N. J. van Druten, D. S. Durfee, D. M. Kurn,
and W. Ketterle, Phys. Rev. Lett. {\bf 75}, 3969 (1995).

\bibitem{gora}P. O. Fedichev, Y. Kagan, G. V. Shlyapnikov, and
J. T. M. Walraven, Phys. Rev. Lett. {\bf 77}, 2913 (1996); J. L.
Bohn and P. S. Julienne, Phys. Rev. A {\bf 56}, 1486 (1997).

\bibitem{verhaar}A. J. Moerdijk, B. J. Verhaar,
and T. M. Nagtegaal, Phys. Rev. A {\bf 53}, 4343 (1996).

\bibitem{eite}E. Tiesinga, A. J. Moerdijk, B. J. Verhaar,
and H. T. C. Stoof, Phys. Rev. A {\bf 46}, R1167 (1992); E.
Tiesinga, B. J. Verhaar, and H. T. C. Stoof, {\it ibid.} {\bf 47},
4114 (1993); J. M. Vogles, C. C. Tsai, R. S. Freeland, S. J. J. M.
F. Kokkelmans, B. J. Verhaar, and D. J. Heinzen, Phys. Rev. A {\bf
56}, R1067 (1997).

\bibitem{marinescu}M. Marinescu and L. You,
Phys. Rev. Lett {\bf 81}, 4596 (1998).

\bibitem{liyou}L. You, and M. Marinescu,
Phys. Rev. A {\bf 60}, 2324 (1999).

\bibitem{huangb}K. Huang, {\it Statistical Mechanics},
(Wiley, NY. 1987).

\bibitem{mm}M. Marinescu and A. Dalgarno, Phys. Rev. A {\bf 52}, 311 (1995).

\bibitem{wu}T. Y. Wu and T. Ohmura, {\it Quantum Theory of
Scattering}, (Prentice-Hall, Englewood, N.J., 1962), p9.

\bibitem{jackson}J. D. Jackson, {\it Classical Electrodynamics},
3rd edition, (John Wiley and Sons, New York, 1999), p149.

\bibitem{greene} C. H. Greene, (private communication).

\bibitem{qdt}M. J. Seaton, Proc. Phys. Soc. London {\bf 88} 801
(1966); U. Fano, J. Opt. Soc. Am. {\bf 65}, 979 (1975);
 C. Greene, U. Fano, G. Strinati, Phys. Rev. A {\bf 19},
1485 (1979); C. H. Greene, A. R. P. Rau, and U. Fano, Phys. Rev. A
{\bf 26}, 2441 (1982).

\bibitem{bogao}B. Gao, Phys. Rev. A {\bf 58}, 1728 (1998);
 Phys. Rev. A {\bf 59}, 2778 (1999).

\bibitem{cavagnero} M. J. Cavagnero, Phys. Rev. A {\bf 50}, 2841
(1994).

\bibitem{fano}U. Fano and A. R. P. Rau, {\it Atomic Collisions and
Spectra}, (Academic, 1986).

\bibitem{bogao2001} B. Gao,  Phys. Rev. A {\bf 62}, 050702 (2000).

\end{references}
\end{document}